\documentclass[12pt,preprint]{aastex}

\slugcomment{To Appear in the Astronomical Journal}

\shorttitle{C\ NICMOS Ultra Deep Field }
\shortauthors{Thompson et al.}

\begin{document}

\title{THE NICMOS ULTRA DEEP FIELD: OBSERVATIONS,\\
    DATA REDUCTION, AND GALAXY PHOTOMETRY}

\author{Rodger I. Thompson}
\affil{Steward Observatory, University of Arizona,
    Tucson, AZ 85721}

\author{Garth Illingworth, Rychard Bouwens}
\affil{Astronomy Department, University of California, Santa Cruz, CA 95064}

\author{Mark Dickinson}
\affil{National Optical Astronomy Observatories, Tucson, AZ 85719}

\author{Daniel Eisenstein, Xiaohui Fan}
\affil{Steward Observatory, University of Arizona,
    Tucson, AZ 85721}

\author{Marijn Franx}
\affil{Leiden Observatory, Postbus 9513, 2300 RA Leiden, Netherlands}

\author{Adam Riess}
\affil{Space Telescope Science Institute, Baltimore, MD 21218}

\author{Marcia J. Rieke,Glenn Schneider, Elizabeth Stobie}
\affil{Steward Observatory, University of Arizona,
    Tucson, AZ 85721}

\author{Sune Toft, Pieter VanDokkum}
\affil{Department of Astronomy, Yale University, New Haven, CT 06520}

\begin{abstract}

This paper describes the observations and data reduction techniques for 
the version 2.0 images and catalog of the NICMOS Ultra Deep Field 
Treasury program. All sources discussed in this paper are based on 
detections in the combined NICMOS F110W and F160W bands only. The 
NICMOS images are drizzled to 0.09/arcsecond pixels and aligned to 
the ACS UDF F850LP image which was rebinned to the same pixel scale. 
These form the NICMOS version 2.0 UDF images. The catalog sources are chosen 
with a conservative detection limit to avoid the inclusion of numerous spurious 
sources. The catalog contains 1293 objects in the $144 \times 144\arcsec$ 
NICMOS subfield of the UDF. The $5\sigma$  signal to noise level is an 
average $0.6\arcsec$ diameter aperture AB magnitude of ~27.7 at 1.1 and 
1.6 microns. The catalog sources, listed in order of right ascension, satisfy 
a minimum signal to noise criterion of 1.4$\sigma$ in at least 7 contiguous 
pixels of the combined F110W and F160W image.
  
\end{abstract}


\section{Introduction}

The NICMOS UDF Treasury observations were designed to complement and enhance 
the ACS optical UDF observations.  They provide an extension in wavelength 
to $1.6 \micron$ and provide two additional bands which extend the rest band 
energy and morphology measurements to longer wavelengths.  They also provide 
the potential for viewing objects at redshifts beyond 7.5 
where Lyman line and continuum absorption quench the flux in the ACS bands.  
The additional wavelength coverage also helps distinguish between the influences 
of age, metallicity and extinction.  As with previous deep field catalog publications
(\citet{wil96}, \citet{thm99} and \citet{beck04}) this paper is intended primarily 
as a description of the observations, data analysis and source photometric properties 
rather than a scientific evaluation of the implications of the observations.

Due to the relatively small field of the NICMOS camera 3 used in this program ($51\arcsec 
\times 51\arcsec$), only a subsection ($144\arcsec \times 144\arcsec$) of the optical UDF was 
covered.  This was done with a 3x3 tiling of the NICMOS images.  The NICMOS images extend a 
few arc seconds beyond this subsection but at a significantly  decreased signal to noise.  
All of the individual processed NICMOS images are available in the HST treasury archive 
(MAST\footnote{http://archive.stsci.edu/prepds/udf/udf\_hlsp.html}) as are the raw images in the main HST archive.

The primary purpose of this paper is to provide a very detailed account of the data
reduction steps used to produce the treasury images and catalog stored in MAST 
 so that users are aware
of the pedigree of the data and can reproduce the analysis if they wish.  Other users
may wish to alter the reduction steps if they prefer other choices than the ones made
here.  Finally we wish to convey the attributes and limitations of the catalog of sources.
Users in particular who wish to extract the faintest sources from the data will want
to use more aggressive extraction techniques than we have utilized in producing the 
treasury catalog.

\section{Observations} \label{s-obs}

The NICMOS observations in the UDF (HST GO/9803) are centered on the 
position 3\fh 32\fm 39.0\fs, -27
\arcdeg 47\arcmin 29.1\arcsec (J2000) at approximately the center of the ACS UDF 
observations (see image at http://www.stsci.edu/hst/udf/parameters/\#Pointing).  
The images lie in a 3x3 grid with centers separated by 45 arc seconds.  The 
grid centers are dithered by 5 arc seconds in a 4x4 square pattern tilted at 
22.5 degrees to the x axis of the images to reduce the effect of 
intra pixel sensitivity variations.  The tilt of the pattern also 
produces dithers that have fractional pixel offsets in the detector array.

The NICMOS UDF program consists of 144 orbits broken into two epochs: i) August 30, 
2003 to September 14, 2003 and ii) November 2, 2003 to November 27, 2003. 
The scheduled start of observations in mid August 2003 was delayed by a safing 
of the NICMOS Cooling System (NCS) just prior to the beginning of observations. 
All reference files such as flats and darks were obtained subsequent to the safing 
event. The epochs are separated to enable the detection of SN Ia candidates and 
to provide enough SAA crossing free orbits for high sensitivity. The images in 
the two epochs are rotated by 90 degrees to accommodate the roll restrictions of the 
spacecraft.  In the first epoch the y axis of the NICMOS camera 3 was oriented 
40.925 degrees east of north, and in the second epoch 130.925 degrees east of north.  
These orientations are the same as the two ACS orientations in an effort 
to align the NICMOS and ACS images without the need for a rotation.  The 
drizzled ACS images in MAST, however, were produced with the standard north 
up and east left orientation.  The version 2 NICMOS treasury images supplied
to MAST are also oriented with north up unlike the version 1 NICMOS image.
The NICMOS version 2 images have been aligned with the ACS version 1 image
so that researchers can utilize the complete set of ACS and NICMOS images
without realignment or rotation, see \S~\ref{ss-dp}. 

\begin{figure}
\plotone{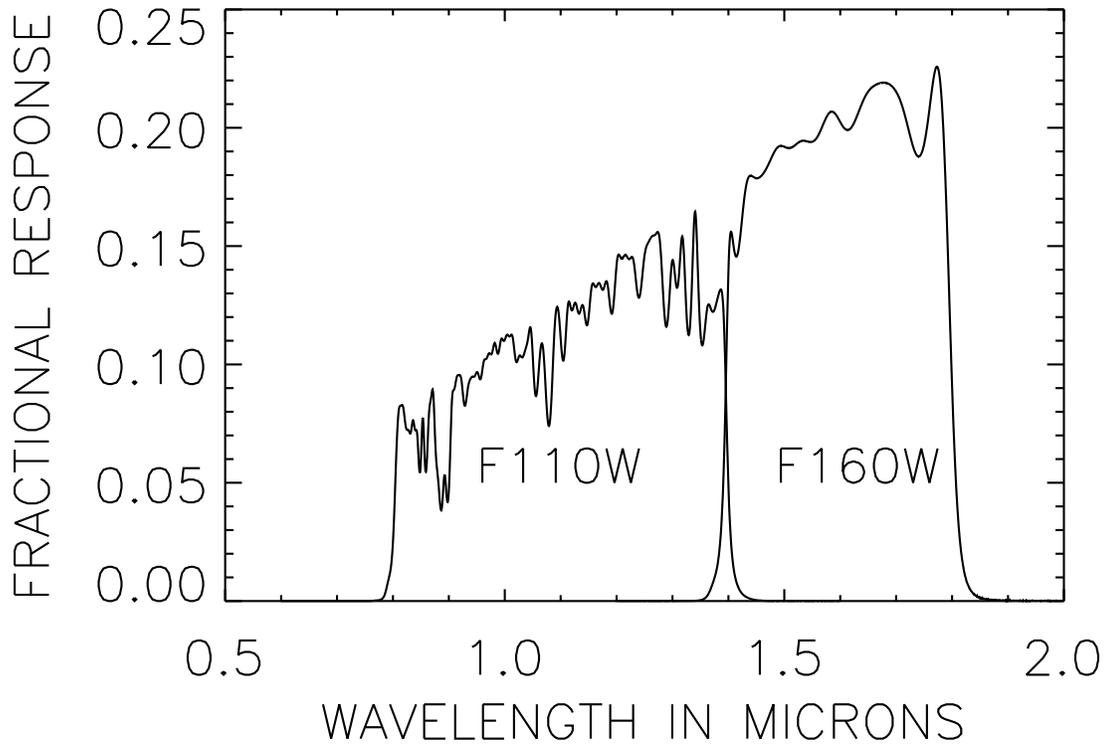}
\caption{The total response functions for the NICMOS F160W and 
F110W filters. $100\%$ response is equal to 1.0 on this plot.}
\label{fig-qe}
\end{figure}

Each orbit contains a F110W and F160W band SPARS64, NSAMP = 24, 1344 second integration. 
Figure~\ref{fig-qe} shows the total response of the two bands, including the detector
quantum efficiency. The F110W integration always preceded the F160W integration. Three orbits 
comprise a single visit with the location of the image specified by a POSTARG from the 
central reference position.  It is possible that the F160W images from the first orbit 
of a visit may have received a small amount of earthshine in the last readout of 
the NSAMP = 24 readout sequence.  
This is due to the guide star acquisition time on the first orbit of the visit which takes 
longer than the guide star reacquisition in the subsequent 2 orbits of the visit.  
No correction has been made in the Treasury images for this effect. The three by three grid 
was completed in three visits with each visit representing a 3 image strip stepped in the 
detector x direction.  The visits were ordered to finish each 3 by 3 grid before starting 
another and always in the same order.  This was done to maximize the time coverage for a 
supernova event, should one occur. After the initial delay due to the NCS safing all of the 
visits occurred in their expected order.

Subsequent analysis of the images indicated that the second half images were not an 
exact repeat of the first half images.  In the second half 3 of the visits replaced 
a center image with an image on the right hand side of the 3 image strip.  
Tables~\ref{tb-pos1} and \ref{tb-pos2} list the visits with T, C, B indicating 
top, center and bottom 
for the epoch 1 visits and L, C, and R indicating left, center and right for the 2nd
epoch visits.  These refer to left, right, top and bottom of the NICMOS images in 
their standard x and y orientations.

\section{Data Reduction}

Previous papers \citep{thm99,thm01,thm03} described details of the data reduction 
procedures for deep field NICMOS observations.  Some of the procedures required 
revision for post NCS installation data.  These changes and the provision of the 
reduced images as a public Treasury Program data products warrants a detailed 
description of the data reduction procedures even though some aspects have been 
covered in the publications cited above.  All of the procedures utilize the 
commercial Interactive Data Language (IDL) \footnote{IDL is a registered trademark of
Research Systems Incorporated, a Kodak company.} software and the IDL based Fits List 
Calculator (FLC) \citep{lyt99} software.  These procedures were 
developed primarily by RIT and ES for previous NICMOS observations.

\subsection{Basic Reduction}

The initial steps in the data reduction take advantage of the nondestructive readout 
capability of the NICMOS detectors.  NICMOS detectors are read out at specified intervals 
without erasing or altering the image.  This ability allows several data reduction advantages 
not shared by CCD detectors.  The SPARS64 read pattern with 24 samples provides readouts 
evenly spaced by 63.998 seconds after the first 2 samples that are spaced by 0.303 seconds.  
All of the images in the NICMOS UDF observations were taken in this mode.  The steps 
described in this section are done automatically in batch processing with no 
interaction.  This is roughly equivalent to the STScI pipeline processing.

\subsubsection{First Read Subtraction}

The first step in the data reduction is the subtraction of the image obtained 
in the first read from all subsequent reads.  This step eliminates the KTC 
noise that is present in each of the individual photodiodes at the beginning of 
an integration.  The number of reads carried through in the final processing is 
then 23 reads rather than 24 after this step. 

\subsubsection{Dark Current Subtraction}

After the first read subtraction the dark current image is subtracted from each of the 
reads.  This is a very important step as the NICMOS detectors have dark current images with 
very significant structure.  This structure is larger in magnitude than the signal from most 
of the galaxies in the image.  The dark images are constructed from integrations in exactly 
the same mode as the observations but with the cold blank filter in place.  This step differs 
from the STScI pipeline that uses ``synthetic darks'' calculated from parameters developed 
during the operation of NICMOS \citep{mob04a}.

\subsubsection{The UDF NICMOS Dark}

The NICMOS UDF dark is a median dark image obtained from dark integrations taken during the 
earth occultation period in each of the orbits assigned to the NICMOS UDF program.  
Operational constraints prevented dark integrations on 2 orbits but the remaining 142 dark 
integrations were used to construct the median images.  There is a median dark image for each 
read constructed from the medians of all of the dark images for that particular read.

Between visit 34 and visit 35 of the 48 visits in the NICMOS UDF program, the temperature set 
point on the NCS was reduced by 0.1 K to compensate for the warmer conditions encountered 
during the period when the earth's orbit is closest to the sun.  There was concern that this 
set point change would alter the nature of the NICMOS darks since it is known that the darks 
are temperature sensitive.  Comparison of a median of the darks taken before the set point 
change with the median darks taken after the set point change did not reveal any differences 
above the noise level in the observations.  The NICMOS UDF dark was therefore constructed 
from the median of all of the darks before and after the set point change.

\subsubsection{Warm Pixels}

In the NCS era the NICMOS detectors operate at a significantly warmer 
temperature than the previous operation with solid nitrogen cryogen.  
There are several advantages to the warmer detector temperature but a 
disadvantage is an increase in the number of ``hot'' and ``warm'' pixels.  
Hot pixels are pixels with a dark current high enough to reach the 
nonlinear response region in a 1000 second integration.  These pixels 
are included in the list of bad pixels described in \S~\ref{ss-badpix}.  
More difficult cases are the warm pixels that have elevated dark 
current but not elevated enough to become nonlinear in a normal integration.  
These pixels can be corrected through the dark image subtraction for 
most types of integrations. The degree of elevation, however, is very 
temperature dependent, and will vary over the normal range of temperature 
variations inherent in the NCS operations.  The elevated dark current is in 
many cases much larger than the signal encountered in faint galaxies.  

The warm pixels contribute noise in two ways.  First is the normal 
Poisson statistics inherent in a steady dark current and the second 
is incomplete removal when the dark current varies from the median dark 
current measured by the NICMOS UDF dark.  The presence of these 
varying warm pixels altered the previous data reductions in two ways.  
The first was the introduction of a more aggressive bad pixel list for 
the UDF observations and the second was the introduction of post 
processing procedures to detect the presence of warm pixel signals 
and to separate them from true sources.  These procedures are described 
in \S~\ref{ss-wpc}. 

\subsubsection{Linearity Correction}

During thermal vacuum testing prior to launch and in cycle 7 the 
linearity of each pixel was measured.  The point a which nonlinearity
set in and coefficients for a polynomial fit after that point were
determined for every pixel.  In the linearity correction stage
the signal of every pixel is checked to see if it is in the nonlinear range.  
If it is above the linear range its signal is corrected using the 
determined coefficients of the polynomial fit. If it is in the signal 
range that is deemed uncorrectable it is marked as saturated and only 
the reads occurring before saturation are used in the analysis. New
coefficients are being determined for post NCS operation. However, none of 
the galaxies in the UDF reached signal levels requiring correction therefore 
linearity is not an issue in this analysis. 

\subsubsection{Cosmic Ray Removal}

A cosmic ray event produces a sharp jump in signal intensity in the first 
readout after the event.  Most events do not saturate the pixel and 
subsequent readouts continue to monitor the incident flux.  At this point 
in the analysis the readouts are stored as delta signal levels 
between each readout.  The signal ``ramp'' is reconstructed by adding the 
deltas together. The first step in cosmic ray detection is a linear fit to 
the signal ramp, which will be a poor fit to the data if there is a cosmic 
ray jump.  The residuals to the fit will be increasingly negative with a 
sharp transition to positive after the event. The cosmic ray detection 
procedure looks for the negative to positive transition as a 
signature of a cosmic ray hit.  If it detects a residual transition above 
the level expected from noise it removes the delta signal between the two
readouts before and after the event, recalculates the signal ramp and 
fits a new linear solution.

The cosmic ray procedure rechecks the ramp to see if there was another cosmic ray hit 
and removes the proper delta signal if one is detected.  If there is still a 
detectable cosmic ray signature after the second refit the pixel is marked as bad 
and no further correction attempt is made.  If the signal is saturated after the 
cosmic ray hit only the signal obtained before the event is used in the analysis.  
The final recorded signal is the value of the slope of the linear fit to the 
signal ramp in adus per second.  Cosmic ray hits that occur in the 0.3 seconds 
between the first and second read are detected as fits that do not intercept 
zero, recalling that the first read is subtracted  from second so that the 
second read is the first point in the ramp.  These hits do not affect the 
calculated slope but are marked as cosmic ray hits in the data quality array 
discussed later.

\subsubsection{Quadrant Bias Correction}

Each quadrant of the NICMOS detectors has a separate output amplifier to transmit 
the analog signal to an A/D converter.  This was done to prevent the loss of an 
entire detector array if there was a failure of an output amplifier.  As a side effect 
of this design decision a small DC bias offset can occur between the 4 
detector quadrants.  Although the offset is small, the effect is significant 
relative to the faint galaxy signals in the UDF and can cause unreliable outputs 
during the drizzle process.  Since there is significant sky background from 
the zodiacal light it is difficult to determine the offset amount from a 
simple inspection of the image.  

The quadrant bias is removed via a procedure based on the bias removal 
procedure developed by 
Mark Dickinson as part of the STScI NICMOS team \citep{mob04b}.  The procedure utilizes the 
flat field imprint produced on the DC signal by the flat field correction process.  Any flat 
DC bias will be modulated by the variations in the flat field.  The process successively 
subtracts a DC bias from each quadrant before it is flat fielded, applies the flat field and 
then picks the bias subtraction that produces the minimum variation in the quadrant.  The 
variation in the quadrant signal is measured by a gaussian fitting to the histogram of 
pixel values in the images. To avoid any residual corner glow from the amplifiers 
or other quadrant boundary anomalies only the quadrant region that is at least 
20 pixels from the quadrant edges 
is used to determine the quadrant bias. Bad pixels are also masked out to prevent them from
dominating the variation signal.  The output of variations from each bias correction is
fit by both a second order polynomial and by 5 point smoothing of the output.  
In the cases encountered in the UDF images they are essentially identical.  The minimum 
variation bias correction is selected as the minimum of the smoothed output.

\begin{figure}
\plotone{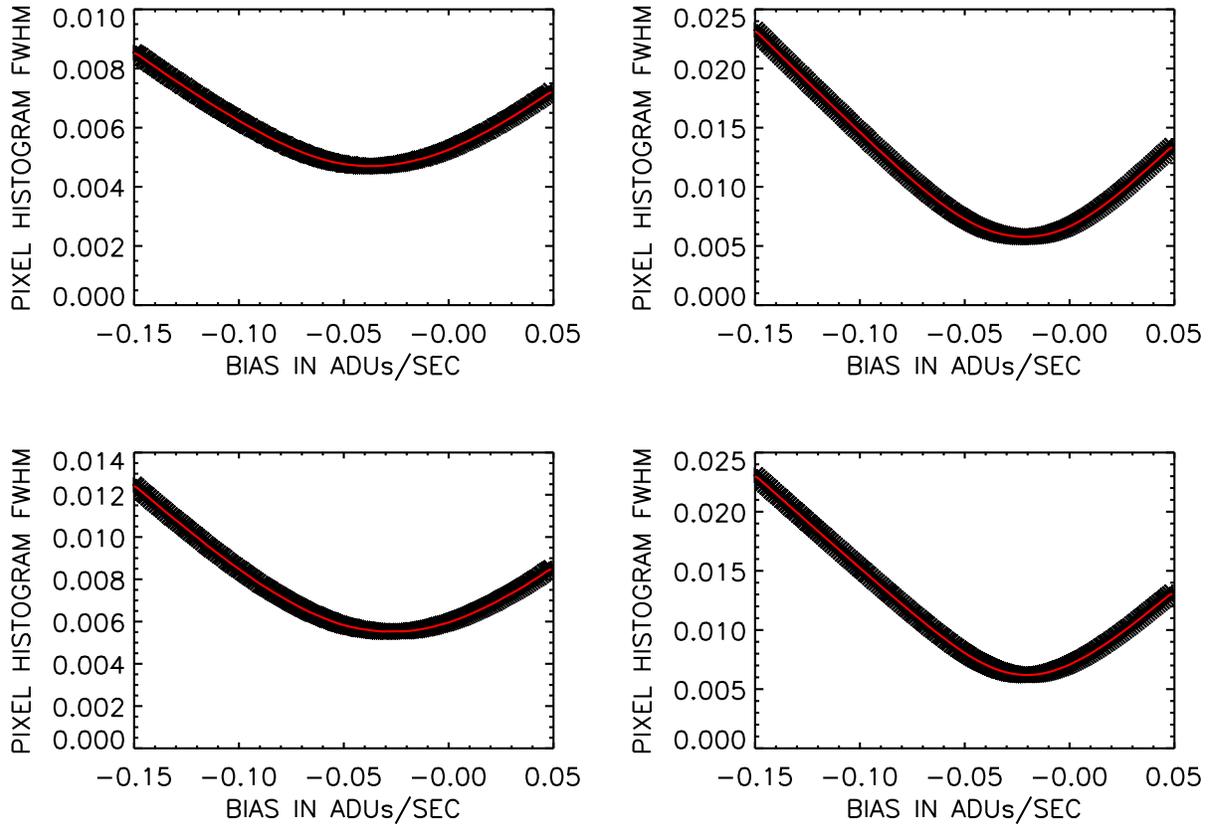}
\caption{The four panels show the quadrant bias corrections in adus per second that
produces the minimum variation due to flat field imprint for one of the
NICMOS images.  These values are taken as
the appropriate quadrant bias correction. The red line shows the polynomial fit to
the output and the thick black line the smoothed fit to the output.  In all cases they completely
overlap.}
\label{fig-bias}
\end{figure}

Both positive and negative bias are subtracted as the bias can have either a positive or 
negative value.  The bias subtraction used biases between -0.15 and 0.2 adus per second
incremented in 0.001 adus per second.
The procedure returns a warning if any bias corrections do not find a minimum in the provided
range of biases.  All of the UDF quadrant images had minimums within the range of biases in 
the procedure.  An example of the gaussian width versus subtracted bias is shown in 
Figure.~\ref{fig-bias}. Visual inspection of the images before and after background subtraction
confirmed that there were no detectable remaining quadrant bias offsets.  
The procedure would be unnecessarily time consuming for images where 
the objects were significantly brighter than the offsets and might not work in images where 
the width of the pixel signal histograms are dominated by source variations rather than 
noise.  Neither is the case for the UDF.  In reality the IDL code for this procedure 
actually performs both the flat fielding and bad pixel correction.  These procedures,
however, are discussed individually in the following sections.

\subsubsection{Flat Fielding}

NICMOS flat fields are created internally.  The ``beam steering mirror'' internal 
to the instrument lies at a optical pupil and is used to correct the spherical 
aberration of the HST primary.  It can be illuminated from behind and the 
reflective coating of the mirror was adjusted to be about $0.01\%$ transmitting 
producing an illuminated pupil for flat fielding.  Flat fields are produced at 
regular intervals during operation in all filters of each camera.  We used F110W 
and F160W camera 3 flat field observations created on Sept. 9, 2003 from 
proposal 9640.  The flat fields are analyzed in the identical manner as described 
in the preceding steps.  The STScI reference flat fields were not used in this 
analysis because they were based on flat fields observed previous to the NCS 
safing event.

One of the effects that the flat field corrects is a slight vignetting along the 
lower edge of camera 3.  For two reasons this correction was not effective in 
the UDF fields.  First, the net effect has two components, vignetting of the 
incoming astronomical flux and emission from the vignetting component which is 
thought to be the edge of the mount for the field division mirror for camera 3.  
For bright sources the vignetting is the dominant effect and the flat field 
properly corrects the field.  For very faint images, such as the UDF, emission 
can be a significant component which varies due to the natural temperature 
variations in the aft shroud. Second, variations in geometry due to temperature 
changes can affect the degree of vignetting.  Again the effect is slight for 
bright sources but can be significant for the UDF signal levels.  For 
these reasons the lower 20 rows of all UDF images were masked off in the drizzle
procedure described in section~\ref{ss-dp}.

\subsubsection{Bad Pixel Correction} \label{ss-badpix}

Bad pixels are defined as pixels with quantum efficiencies less than $10\%$ of 
the average QE or with dark currents high enough to reach nonlinear signal 
levels in 1000 seconds or less. In the post NCS era the list of bad pixels 
has increased over the cycle 7 listing due to the higher temperature of the 
detector creating more high dark current pixels.  Pixels which satisfy neither 
criterion but are highly variable in their 
dark current were also added to the list.  All bad pixel signals are replaced with the 
median of the eight pixels surrounding them.  In the case of adjacent bad pixels that 
number is reduced by the number of adjoining bad pixels.  All bad pixels are listed 
in the data quality array which is an extension of the image or SCI array.  
Table~\ref{tb-dqc} gives the decimal codes for each of the steps described above.  
They are each a single, different, binary bit, so each combination of actions 
performed on a pixel has a unique output code.  Note that only a few of the 16 bits 
available for pixel actions are used in this analysis.  A full set of data quality 
codes can be found in the NICMOS Handbook but only the ones listed here are used 
in the NICMOS UDF Treasury data.  Even for the codes used they may in many cases 
differ from the codes returned by the STScI pipeline analysis.  As an example, the 
lists of bad pixels differ between the pipeline analysis and the analysis described 
here.  Note that the data quality extensions only exist for the individual NICMOS 
UDF images.  The drizzle procedure does not preserve these codes since many input 
pixels contribute to a single drizzled output pixel.

\subsection{SSA Persistence Correction}

The program planners at STScI were careful to schedule the NICMOS UDF observations 
in orbits that were not impacted by SAA passages.  None of the NICMOS UDF images 
required any correction to remove SAA persistence.

\subsection{Earthshine Detection}

As mentioned in \S~\ref{s-obs} the last read of the first orbit in a visit may have 
encountered increased earthshine due to the longer period of delay in a guide star 
acquisition rather than reacquisition.  This would only affect the F160W images as 
the readout sequence was always F110W and then F160W. We tested for this effect by 
plotting the median of the delta increase in the F160W images as a function of 
readout. In a few cases we saw a detectable rise in the last readout of the 
first orbit in a visit.  The effect appeared minor enough that no correction 
was attempted.

\subsection{Background Subtraction}

The primary background source in the UDF is zodiacal emission which is relatively 
uniform across the small UDF field of view.  A median image of all of the images in 
the F110W and the F160W filter determines the background for that filter.  
The background is simply subtracted from each image in the proper filter.  
The median image is extremely smooth with no indication of any residual 
source structure.  This is similar to the results in the NHDF \citet{thm99} where 
the spacing between images was much smaller than in the UDF. Since the 
zodiacal backgrounds can be time dependent the first epoch and second epoch 
background subtractions were done independently.

\subsection{Residual Bias Correction}

The removal of quadrant biases and the background subtraction should result in an 
image that has a median value of nearly zero since most of the pixels in the image 
are not in sources that are above the noise level. The minimums of the quadrant bias
curves in Figure~\ref{fig-bias}, however, are rather broad.  To compensate for this
each quadrant of every image was set to zero bias.  The bias was determined from 
the median of the portion of the quadrants which
is 40 rows and columns away from the edge to avoid any contamination from residual
corner glow or uncorrected vignetting. Any detected bias was subtracted from the 
entire quadrant to produce a zero bias image.  Inspection by eye of the zero bias
images did not find any detectable quadrant offsets.

\subsection{Warm Pixel Correction} \label{ss-wpc}

A new effect encountered in the NICMOS UDF images is warm pixel variation.  At 
the higher operating temperature of the NCS cooling system, the NICMOS detectors 
have more warm pixels than during the cycle 7 operation with solid nitrogen cryogen.  
A warm pixel is defined as a pixel with an elevated dark current which is not high 
enough to be declared a bad pixel by the criteria defined in \S~\ref{ss-badpix}. 
For most applications warm pixels are adequately corrected by dark subtraction 
but for the UDF they present two problems.  The first problem is that their dark 
current is temperature sensitive and the NCS has slight temperature variations 
within an orbital cycle and within a 24 hr day due to power cycles in HST operation.  
The variation is small but is significant relative to faint UDF sources.  The 
second problem is that the Poisson noise of the signal at the end of an integration 
is also significant relative to a UDF source.  The individual images have single 
or sometimes double pixels with signal levels of either positive or negative 
high contrast relative to the surrounding pixels from the warm pixel effect.

Warm pixels are identified by their contrast with neighboring pixels.  The PSF of a point
source centered on a pixel provides a maximum contrast for a real source.  Pixels with 
contrasts significantly greater than this are due to warm pixels.  The contrast was 
computed for each pixel compared to the 8 pixels adjacent to the pixel.  A pixel is set
equal to the median of the 8 pixels if 3 conditions are met; i) its value is higher 
than the 3$\sigma$ noise value in the image, ii) its value is higher relative to 
the median of the adjacent pixels than a preset contrast value and iii) the value of 
the surrounding median is less than a preset number of standard deviations.  The last
condition prevents peak clipping on point objects.  The contrast values were set to 
6.0 for the F160W image and 8.0 for the F110W image which has a narrower PSF.  The 
standard deviation limits were set to 2.5 for the F160W images and 2.0 for the F110W
images.  Any corrected pixel has the bad pixel flag set in the DQ image extension 
so it can be identified.  

\subsection{Drizzle Procedure} \label{ss-dp}

The treasury mosaic images were produced using the DRIZZLE procedure with 
context images \citep{fru02}. The offsets for the NICMOS images were 
determined by registering the NICMOS F110W images onto the ACS F850LP UDF 
image in the HST archive which has a north up orientation and $0.03 \arcsec$ pixel
scale. The significant overlap between the two filters greatly reduces any errors due to
color dependent morphology, however, see \S~\ref{sss-psf} for an assessment of the
accuracy of the alignment.  The NICMOS F160W images which always immediately followed
the F110W images in an orbit were assumed to have the same offset as the F110W images
preceding them.  The ACS image was reduced to $0.09 \arcsec$ pixels by a simple 3x3 
pixel addition of the image.  Individual NICMOS F110W images were then produced with
a drizzle PIXFRAC parameter of 0.6 and a SCALE parameter of $0.09/0.202863$ to 
produce $0.09\arcsec$ output pixels. The denominator in the scale factor is the 
pixel size of the distortion corrected NICMOS pixel. 
The geometric distortion coefficients \cite{berg04} are given in Table~\ref{tb-dis}.
These coefficients are the constants for a cubic distortion correction of the form

\begin{equation}
xdist = a1 + a2\times x + a3 \times y +a5 \times x \times y + a6 \times y^2 + a7 \times x^3 + a8 \times x^2 \times y + a9 \times x \times y^3
\end{equation}

\noindent and an identical equation in b coefficients for the y position that governs
the placement of the pixels.  Compared to other HST instruments the correction is
relatively small.  The main component is the difference in plate scale between 
the x and y directions due to a slight tilt in the camera 3 focal plane relative 
to the plane of the detector. The tilt is due to the curvature of the focal plane.
Each $0.09 \arcsec$ NICMOS was rotated to a north up orientation using the ORIENTAT
value in the image header.

A three step process provided the positions of each F110W image relative 
to the ACS F850LP image.  The first step was to shift the NICMOS images 
to the positions indicated by their World Coordinate System (WCS) position 
in the headers.  The second step was a non-interactive chi squared 
minimization of the differences between the bright objects in the NICMOS 
image and the nearest corresponding bright ACS object.  
The ACS positions were determined with SE in the ABSOLUTE mode with the
threshold set at 0.03 ADUs per second.  Positions in the NIMCOS images 
were also determined with SE in ABSOLUTE mode with the threshold set at
0.01 ADUs per second.  Both of these thresholds are quite bright to insure
a low source count per area.  This made the likelihood of wrong object
matching low.  The shifts were limited to plus or minus 10 pixels in 
the X and Y directions in single pixel steps. The average position shift 
in this step was on the order of 2 to 3 $0.09 \arcsec$ 
pixels.  The final shifts were determined by a similar chi squared minimization
of interactively selected NICMOS objects.  Usually three objects were selected
based on compact size and sufficient signal to noise.  Whenever fully visible
in a NICMOS image, the star near the center of the image was used as one of
the objects.  The more eastern star image appears to have contamination due to 
a faint nearby object visible in the ACS images. Shifts in this step were 
limited to plus or minus 1 pixel in 0.1 pixel steps.  The average position 
adjustment in this third and final stage was 0.2 to 0.3 $0.09 \arcsec$ 
pixels.  These final positions were then used as the input to the drizzle 
procedure. 

\subsubsection{Individual Image Masking} \label{sss-msk}

The F110W and F160W images each have a general mask used in the drizzle process.  The
masks mask out the bottom 20 rows of the image to eliminate the partially vignetted
region at the bottom of camera 3.  They also mask out a portion of the upper right
hand corner of camera 3 where there is an area of rapidly changing quantum efficiency.
In addition they also mask out the known bad pixels and areas where some particles,
termed grot, cover detector pixels.  The large number of dithered images greatly
reduces the impact of the masked areas.  The masks are available in the STScI NICMOS
UDF Treasury version 2.0 archive in MAST.
Several of the images had artifacts such as satellite passage streaks that required
masking.  Individual masks were made for 20 F110W images and 31 F160W images.
Table~\ref{tb-msk} lists the masked images.  The masks are contained in the NICMOS
version 2.0 Treasury submission.

\subsubsection{Cosmic Ray Persistence} \label{sss-crp}

A few of the masks remove spurious objects created by cosmic ray persistence.
If a cosmic ray hit before the start of an integration produces a shower of particles,
cosmic ray persistence can give a resolved source that appears in the first F110W
image after the hit and more weakly in the following F160W image.  This exactly mimics
a high redshift galaxy.  The ACS image has no signal and the F110W-F160W color is blue.
The signature of this spurious event is that the source only appears in one set of
F110W and F160W images.  All of the NICMOS sources that did not have ACS counterparts
were inspected in each image.  Two spurious cosmic ray persistence sources were found
and masked out (see also \S~\ref{ss-igi}).

\subsubsection{Point Spread Functions} \label{sss-psf}

Unlike the deep NICMOS observations in the NHDF, the HST secondary mirror was not
adjusted to bring the camera 3 images into sharp focus.  The photometric gain was
not considered high enough to request the adjustment which would have put the
parallel ACS images significantly out of focus. The PSF at the focal plane of 
camera 3 is therefore broader than the diffraction limited PSF observed by 
NICMOS cameras 1 and 2.  To determine the PSFs of the UDF images two 
measurements were performed.  The first was to measure the PSF of the bright 
star at x = 1897.26, y = 1610.34 with Gaussian fitting.  The second bright star 
at x = 1246.12, y = 1420.42 appears to be double with a faint companion.  
The second measurement involved 42 camera 3 images in F110W
and F160W of the photometric calibration star P330-E taken after the NCS safing
event.  These were part of the Prop. 9995 calibration program of Mark Dickinson.
These images were drizzled in the same manner as the UDF images.  The PSF of the
drizzled F110W and F160W images were measured by the same Gaussian fitting as for
the UDF stellar images.  The results are listed in  Table~\ref{tab-psf} which 
gives the measured major and minor axis FWHM values.  Table~\ref{tab-psf} also 
gives the results of performing the same exercise on synthetic images produced
with the Tiny Tim software \citep{kri04} for the camera 3 focus utilized in the UDF 
observations. The Tiny Tim and calibration star PSFs agree quit well but 
the measured UDF stellar values are between 0.06 and 0.1 arc seconds wider.  
This may reflect the accuracy of the mosaic position calculations.
The widths measured in an independently reduced image (see \S~\ref{ss-igi})
are very similar to our UDF image widths.  Any researcher that requires 
extremely accurate object shapes, such as for weak lensing, may wish to 
go to the original single images for size and shape measurements. Those 
researchers should apply the geometric distortion corrections listed in 
Table~\ref{tb-dis}.

\section{Image Size and Weight}

The full drizzled image does not have a uniform integration time over the image.
In particular the edges of the image have only one integration as opposed to the 
average 16 integrations for the interior of the image.  The full drizzled image
has a size of 3500 by 3500 pixels, the same size and orientation as the ACS images
reduced to $0.09 \arcsec$ pixels. Experience with the NHDF images indicated that 
source extraction in regions with less than half of the
average integration time was not profitable except in special cases
where a particular object near the edge of the total image required analysis.  Users
of the treasury image should be aware that the edges of the image have roughly a
$\sqrt{2}$ lower signal to noise than the central regions.  The exact weight for all
pixels is given by the weight images included in the treasury archive.  There are
some regions where the images overlap that have much higher integration times.  Users
who need a uniform selection criterion for analysis should be aware of these 
differences in weight.

\section{Source Extraction}\label{s-se}

Source extraction in the science image was performed with the source extraction
program SExtractor (SE) version 2.3 \citep{ber96} in the dual image and rms
image mode. The source extraction includes the 4 ACS UDF images as well as
the NICMOS images.  The ACS extractions provided the source reality check 
described in \S~\ref{ss-acs} and are included in the treasury catalog.

\subsection{Noise} \label{ss-noise}

\begin{figure}
\plotone{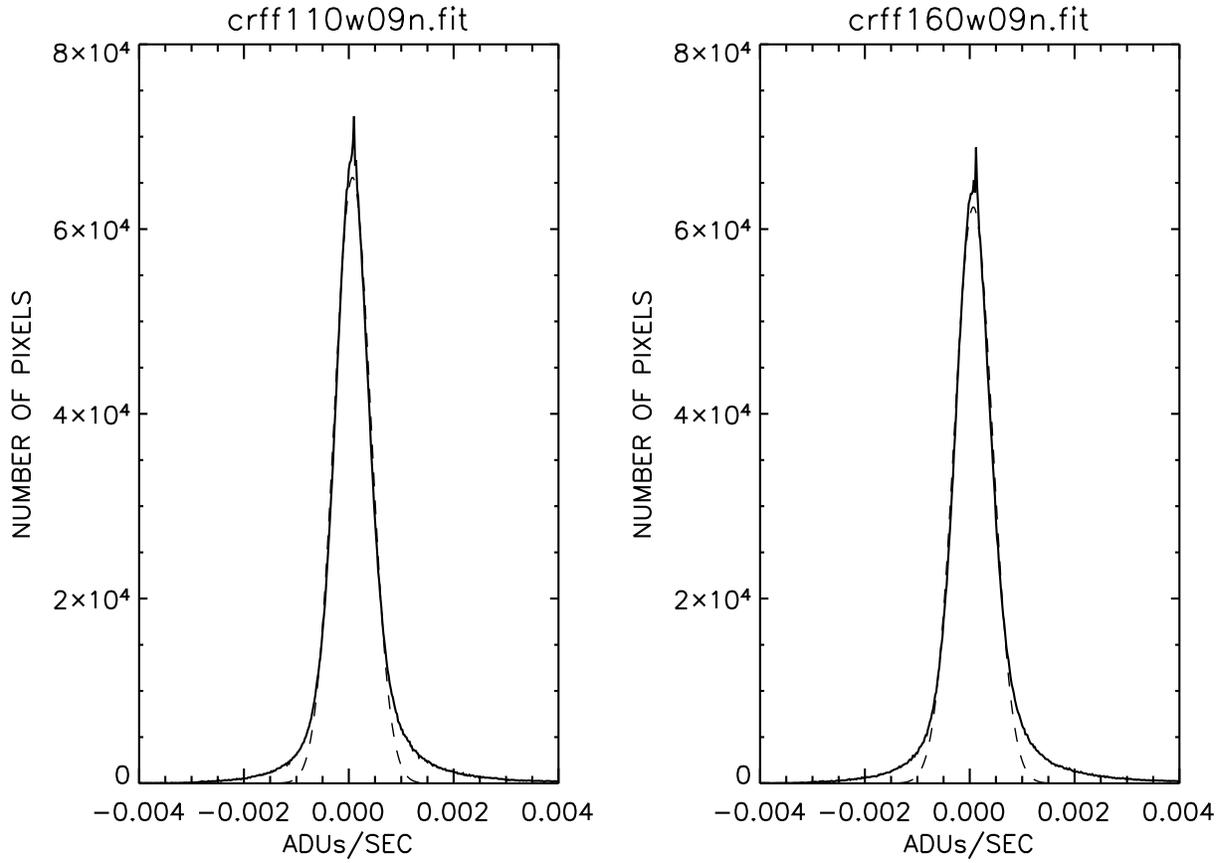}
\caption{Histograms of the pixel values in the F110W (left) and F160W (right) images.
The deviation from the gaussian fit on the positive side is due to the contribution of 
sources.  The width of the gaussian is an indication of the noise in the images.  See
\S~\ref{ss-noise} for a discussion of the pixel noise.}
\label{fig-ns}
\end{figure}

Figure~\ref{fig-ns} shows the histograms of all of the pixel values in the 
F110W and F160W image.  The majority of the pixels are well fit with a gaussian 
centered on zero. The offsets from zero are $~7\times 10^{-5}$ ADUs/second for 
both images. The positive tail deviating from a gaussian is the contribution
from the true sources in the field.  The width of the gaussian fit is an 
indicator of the noise.  The gaussian fit gives noise levels of 
$3.5\times 10^{-4}$ ADUs per second for the F110W image and 
$3.7\times 10^{-4}$ ADUs per second for the F160W image.  This corresponds 
to 0.55 and 0.58 nanojanskys respectively.  This is an underestimate of the 
true noise as is partially indicated by the excess over the gaussian on the 
negative side of the fit.  The drizzle procedure is known to introduce 
correlation to the noise \citep{fru02}. \citet{fru02} give an expression for 
the noise increase factor which gives a factor of 1.8 for the drizzle PIXFRAC 
and SCALE parameters used in the images.  This yields $1\sigma$ noise values 
of 1.0 and 1.2 nanojanskys per pixel.

\begin{figure}
\plotone{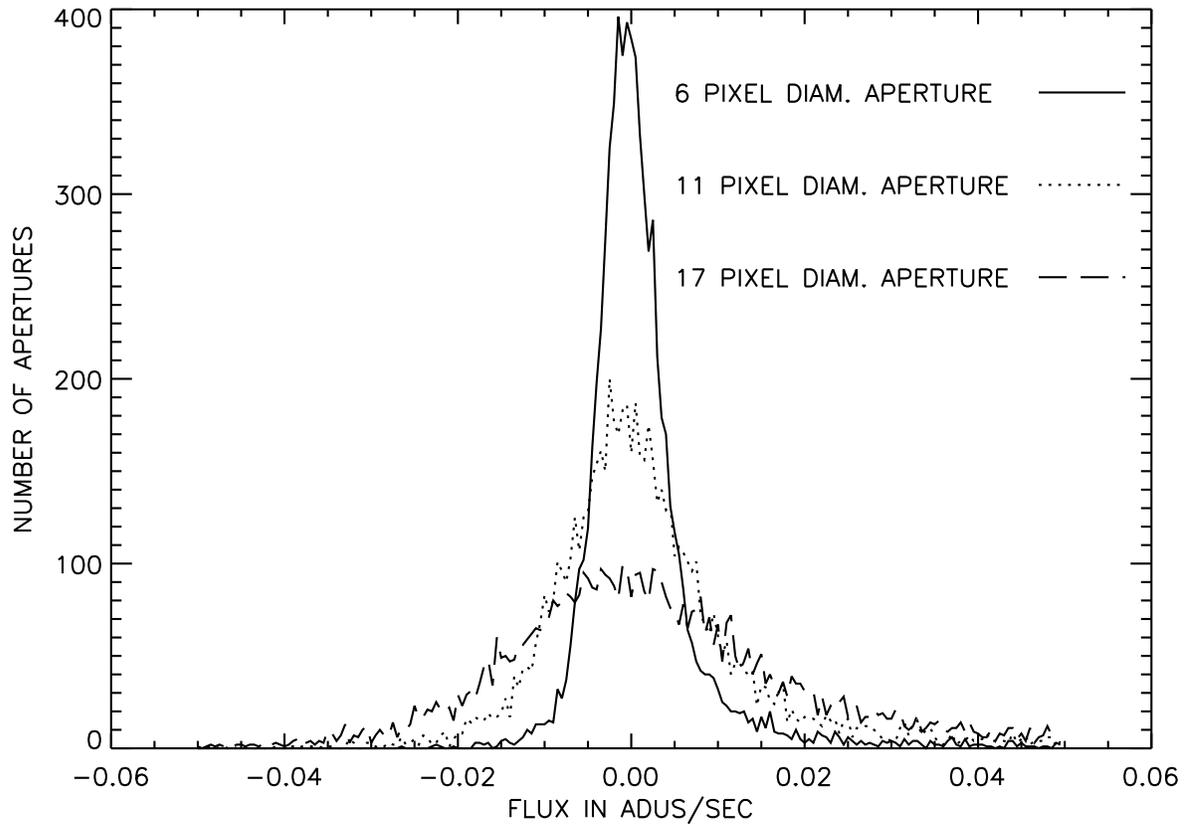}
\caption{Histograms of the flux in a densely packed grid of apertures covering
the UDF F110W image.  The three apertures are 6, 11 and 17 pixels wide. See 
\S~\ref{ss-noise} for a discussion of the aperture noise.}
\label{fig-apn}
\end{figure}

Figure~\ref{fig-apn} shows the histograms of the signals in a densely packed grid
of apertures of the same diameter as the three apertures, 6, 11 and 17 pixels, 
used in the source extraction. The 175x175 grid is regularly spaced on 20 pixel 
centers to provide 30,625 apertures.  Most of these do not contain sources 
but the positive tail in Figure~\ref{fig-apn} indicates apertures with 
positive source flux. The large number of apertures that fall off the 
NICMOS image are not included in the analysis. The histograms are 
roughly Gaussian shaped but only the 6 pixel aperture
histogram has an easily measurable FWHM.  The $1\sigma$ noise derived from the
FWHM is $3.7 \times 10^{-3}$ ADUs per sec for the F110W image. Similar results
were obtained for the F160W image. The expected noise from the 
individual pixel noise described above, including the factor of 1.8, is 
$3.5 \times 10^{-3}$ which is comparable to the measured value.  To the degree
that the aperture noises are truly Gaussian distributed, this indicates
that factor of 1.8 is a reasonable figure to account for the correlated noise
and that the aperture noise is approximately equal to the square root of the
number of pixels in the aperture times the individual pixel noise.  The
measured aperture noise in Janskys from the histogram is $5.8 \times 10^{-9}$
which is equivalent to an AB magnitude of 29.5 for the $0.54\arcsec$ diameter
aperture. The $5\sigma$ AB magnitude is 27.7 which we will take as the 
appropriate value independent of source noise.
In observations of the NHDF-S \citet{lab03} with ISAAC on the VLT
found $1\sigma$ aperture noises of 28.6 and 28.1 for the J and H bands
with a $0.7\arcsec$ diameter aperture which gives $5\sigma$ values of
26.7 and 26.2. 
The minimum number of contiguous pixels for a real source is set to 7 in
the extraction procedure.  The $5 \sigma$ noise for a point source detection
is then the noise in the 7 pixel aperture which gives AB magnitudes
of 30.35 and 30.15 for the F110W and F160W filters respectively. 

\subsection{Detection Image}

SE in the two image mode uses a detection image to determine the position 
and extent of sources.  The individual image source extraction is then 
performed on exactly the same positions and regions determined from the 
detection image.  The SE parameters regarding source geometry such as area 
and ellipticity are determined by the detection image.  The detection image 
for the treasury catalog is the simple sum of the F110W and F160W science 
images.  Even though it is the sum of two images, the detection image has a 
significantly lower signal to noise than any of the ACS UDF images except 
for sources that are extremely red.  Users that are interested in the
NICMOS limits on faint ACS UDF sources should use the ACS images as the 
detection image to perform the source extraction on the NICMOS images.  
On the other hand very red sources may appear only in the NICMOS images.  
Since we wish to provide an infrared catalog we chose to use the NICMOS 
images for extraction.  Note that by combining the two NICMOS images there 
is a bias against the very reddest sources that might only appear in the
F160W image.

\subsection{RMS Image} \label{ss-wi}

SE utilizes a rms image to determine the detection limit of a pixel signal when
operating in the RMS mode used in the treasury version 2.0 catalog.  The
drizzle procedure produces an observation time weight map that measures 
the total integration time for every pixel but the weight map does not 
take into account the large variations in quantum efficiency over the face 
of the NICMOS detector array.  To account for the QE variations, the F110W 
and F160W flat field images were drizzled in exactly 
the same way as the UDF images, utilizing identical masks.  Note that the flat 
field images are the multiplicative flats used in the normal image reductions.
As such they are high where the signal is low and vice versa. All subsequent 
post drizzle procedures that were applied to the science images were also 
applied to the drizzled flat images. Next the median of the drizzled flat 
images was set to one and multiplied by the standard deviation determined in 
\S~\ref{ss-noise}, adjusted by the correlation parameter of 1.8. These images 
were then added together as the square root of the sum of the squares to 
form the rms image for the detection image supplied to SE.  This rms image 
properly represents the differences in QE across the detector and the 
mapping of the differences onto the final science image.  The individual 
F110W and F160W rms images were retained for use in the source extraction 
described below.  The ACS rms images were assumed to be uniform and equal to 
the standard deviations found in \S~\ref{ss-noise} for the individual ACS 
images.  The individual filter rms images form the basis of the magnitude 
and flux errors returned by SE.

\subsubsection{RMS Image Adjustments} \label{sss-ria}

Visual inspection of the detection image and the individual NICMOS drizzled images
revealed small areas where there were clearly higher regions of noise or residual
spurious artifacts.  These are usually regions of ``cross hatched'' noise or small 
irregular regions of a few pixels with boundaries too sharp to be real sources.
Even though areas such as these were masked in the original images, the drizzle
process can produce additional areas due to rebinning process. Rather than 
removing these regions from the images, the rms images were adjusted to guarantee
that they were not used in the source extraction.  The pixel regions with 
adjusted rms values are given in a table supplied with the version 2 submission.
These regions can also be identified in the rms images themselves where they
have been set to 9.999. This procedure preserves the ignored areas 
in the images so that they can be evaluated by other researchers 
if they wish, just as the individual images, before masking, can 
be retrieved from the archive.  Areas outside of the observed regions
were set to very high rms values on the order of 500. to 800.

\subsection{Extraction parameters}

SE's source extraction process is controlled by a configuration file that 
gives the parameters for source extraction.  Some of the configuration 
parameters used in the version 2.0 treasury source extraction are given in 
Table~\ref{tb-separ}. The full configuration files used in the extraction 
are included in the Treasury NICMOS UDF version 2.0 submission to MAST. 
The parameters include a detection threshold of $1.4 \sigma$ and the 
minimum number of contiguous pixels for a true source set to 7.  Note 
that these parameters discriminate against faint point sources.

Although not evident from the documentation, SE runs in a dual rms image 
mode.  In this mode two rms images are supplied in the WEIGHT\_IMAGE 
parameter.  The first image is the rms image for the detection image and 
the second is the rms image for the extraction image which is the UDF image 
in each of the individual filters. The second rms image does not influence 
the source selection but determines the error values returned by SE.
 
The photometric zero points in AB magnitude are 23.41 (F110W) and 23.22 (F160W). 
The extraction parameters were adjusted to produce a clean extraction of sources 
with a goal of no erroneous detections in the high signal to noise regions of the image.  
This produces a relatively conservative catalog given the range of weighting 
over the image. Visual inspection of the image indicates
that there are real sources that have been missed by the extraction process.  More 
aggressive extraction parameters pick up these sources but also begin to find sources
of doubtful reality. (See \S~\ref{s-rel}.) Users who wish to have a more aggressive
extraction can use our parameters as a starting point and adjust them to produce the
required level of extraction.  Our parameters produced 1293 extracted sources.

\section{Catalog Construction}

All of the output values listed in the treasury catalog are outputs
from SE with no editing.  The source order has been rearranged to be 
in order of increasing right ascension.  To maintain correspondence 
with the segmentation images that were returned by SE, the original 
source identification numbers assigned by SE are given in the catalog. 
Ten of the sources detected by SE are not included in the catalog as 
described in \S~\ref{ss-acs} leaving 1283 of the 1293 detected sources. 

Each source has 90 entries, therefore, we do not include the table 
in the paper.  The catalog is available at the STScI MAST site containing 
the NICMOS version 2.0 high level science products.  That catalog is a 
comma separated text file.  Columns 1-2 contain the ID number and the 
ID number of the associated source in the version 1
I band based ACS UDF catalog.  The associated ACS source is the closest source 
to the catalog source.  If there is no ACS source within 0.3 seconds of the 
NICMOS source a 0 entry is made in column 2.  Column 3 gives the 
distance to the associated source in arc seconds and column 4 gives 
ID number of the source in the SE produced segmentation image which is 
also provided in the MAST archive.  If a version 2 catalog of ACS sources 
is produced we will attempt to provide a version 2.1 catalog
with the new ACS source identifications included.

Columns 5 and 6 give the x and y positions in pixels of the source in 
the version 2.0 treasury image. Columns 7 and 8 are the RA and DEC 
positions in degrees while columns 9 and 10 are the RA and DEC position 
in traditional nomenclature. Columns 11-28 give the aperture AB magnitudes 
of the source.  The 3 aperture magnitudes of the ACS F445W band are listed 
first followed by the remaining ACS and NICMOS bands in order of wavelength. 
Columns 29 through 34 list the isophotal magnitudes and columns 35 through
40 list auto magnitudes returned by SE in the same order.

Column 41 lists the number of pixels associated with the source in the SE 
segmentation image.  Columns 42 through 47 list the FWHM in each band. Column 48
lists the position angle of the source returned by SE. Column 49 lists the 
flag value returned by SE. No sources have been removed from the
catalog on the basis of the value of the SE error flag. Columns 50-55 
list the XPEAK, YPEAK, XMIN, YMIN, XMAX,and YMAX values returned by SE.  
The source is contained in a box defined by the minimum and maximum x and y 
values.  Columns 56 and 57 give the ellipticity and elongation of the source 
returned by SE. Column 58 contains the $0.6\arcsec$ diameter aperture AB 
magnitude of the F160W band.

Columns 59-76 contain the aperture fluxes in the same order as the 
aperture magnitudes. The fluxes are in ADU/sec.  The NICMOS gain is 6.5 
electrons per ADU. Columns 77 through 82 list the isophotal fluxes and 
columns 83 through 88 list auto fluxes returned by SE.  Columns 89 and 
90 are the ISOAREA and ISOFAREA values returned by SE.

\subsection{Mini-Catalog} \label{ss-mc}

To be consistent with the ACS UDF submissions we have also constructed a
mini-catalog of the sources, part of which is included in the printed version
of the paper.  The whole catalog is available in the electronic version of
the paper.  The catalog appears in Table~\ref{tab-cat}.  There are some 
differences relative to the ACS catalog available in MAST.  First the catalog is
ordered in RA with the associated ACS source ID and the segmentation ID in 
columns 2 and 3.  The x and y positions followed by the RA and DEC in degrees
are in columns 4-7. Next are the position angle, ellipticity, half radius, FWHM 
and stellarity in columns 8-12.  These are followed by the isophotal 
magnitude,isophotal magnitude error, and signal to noise for the 4 ACS
and 2 NICMOS bands starting with the ACS F435W and ending with the NICMOS
F160W band.  The signal to noise is the ratio of the isophotoal flux to the
isophotal flux error returned by SE. The last entry in the table is the 
value of the error flag returned by SE.  As in the larger table there has 
been no effort to remove sources based on the SE error flag.  Due to space
constraints only the first few columns of the table are present in the
printed version.

\section{Source Reliability} \label{s-rel}

Although the extraction parameters were adjusted conservatively, 
independent assessments of the source reliability are required.  The 
following analyses and tests were performed to judge the reality of 
the catalog sources. 

\subsection{Signal to Noise Values} \label{ss-snv}

\begin{figure}
\plotone{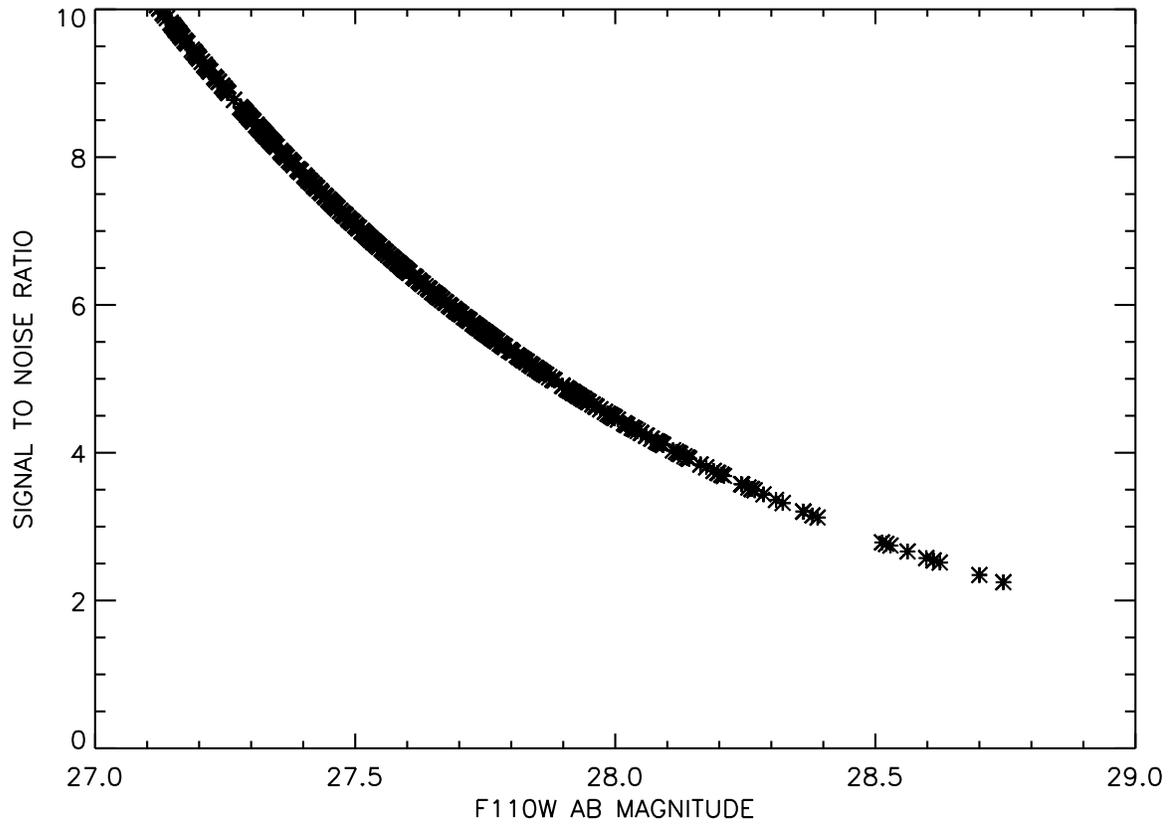}
\caption{The faint end of the isophotal F110W AB magnitude versus signal 
to noise plot where the signal to noise is calculated from the isophotal 
flux and isophotal flux error returned by SE. Each source is marked by an 
asterisk symbol. In general the flux error returned by SE is smaller than
that calculated by other means.  See the discussion in \S~\ref{ss-snv}}
\label{fig-snv}
\end{figure}

The source extraction program, SE, returns flux errors as well as fluxes.
Figure~\ref{fig-snv} shows the measured signal to noise values for the
faint end of the catalog.  The ratio of the isophotal flux to isophotal
flux error is ploted versus isophotal magnitude.  There is significant
scatter in the values as expected.  The average magnitude for a signal
to noise ratio of 5 appears to be around an isophotal magnitude of
28.4 significantly fainter than the aperture test value of 27.7 found in
\S~\ref{ss-noise}.  The details of how SE computes its flux errors
are not immediately obvious so the value of 27.7 will be used.  The
rms images supplied to SE were multiplied by 1.8 for the expected
correlation due to drizzling so that should not be an explanation
for the difference.

\subsection{Negative Image}

As a check on noise induced sources we ran the identical extraction procedure 
on the negative of the original source detection image.  The procedure produced 
no detections from the negative image. This indicates that the number 
of sources produced by noise is very low. 

\subsection{Comparison with the ACS images} \label{ss-acs}

The presence of the much higher signal to noise ACS images provided a 
second test of source reliability.  We checked for catalog sources that 
had ACS F850LP $0.6\arcsec$ aperture diameter AB magnitudes fainter than 29.5. 
The smallest aperture was chosen to minimize flux from overlapping sources.  
We identified 22 sources out of the total 1293 sources that matched that 
criterion.  Two of these sources are in the high redshift source list
of 5 sources described by \citet{bou04} where the Lyman break occurs 
to the red of the ACS cutoff. \citet{bou04} used a more aggressive source
extraction which accounts for the extra three sources in their analysis. 
The remaining 20 sources fell into 3 categories.  The first were legitimate 
sources with ACS F850LP flux below the limit but clearly there under 
visual inspection. There are 10 sources in this category.  The second 
category is sources with clear flux in only the F160W band.  There are 
2 sources in this category. The remaining 8 were sources that appear to be 
noise artifacts by a subjective visual analysis. The 10 objects in the
last two categories do not appear in the catalog even though the 2 sources
with only F160W flux may be real.  

\subsection{Comparison with an Independently Generated Image} \label{ss-igi}

The NICMOS UDF images were independently reduced at STScI by 
Massimo Stiavelli, Bahram Mobasher and Louis Bergeron.  They very 
graciously provided these images for comparison with our reductions.  
Inspection of these images revealed two objects in our images 
that were the result of cosmic ray persistence (See \S~\ref{sss-crp}).  
These objects were masked out in the individual image they appeared in 
for the final drizzle run. One NICMOS object that does not have an ACS 
counterpart, object 937, \citet{bou04} UDF 818-886, does not appear
in the Stiavelli, Mobasher and Bergeron image. Inspection of our 
individual images indicates that no single image contributed the 
majority of the flux as would be expected for a cosmic ray persistence 
event.  The object appears faintly in several of our individual
images in both filters.  To date we have not been able to resolve 
why this discrepancy exists.  Users of the catalog should
be aware of this discrepancy, particularly because this source 
satisfies the criteria for a galaxy at a redshift greater than 7.

\section{Conclusions}

\begin{figure}
\plotone{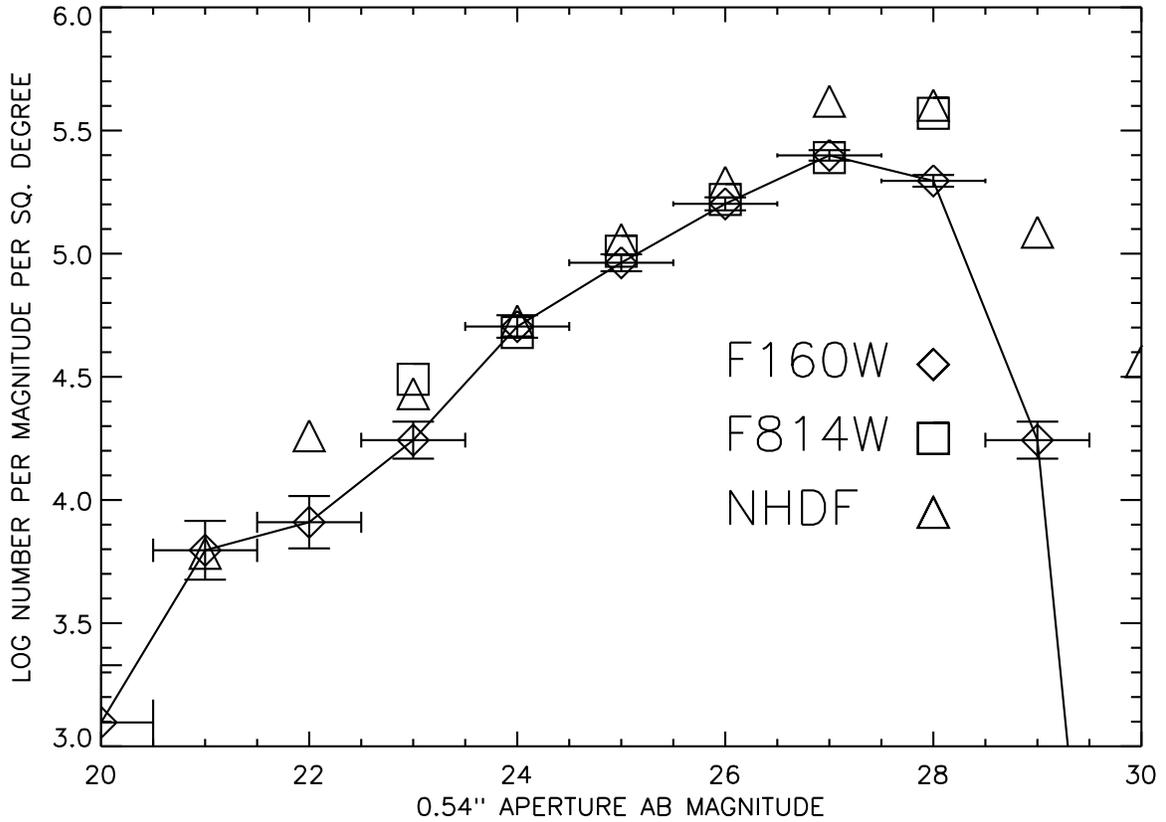}
\caption{The H band (F160W) number magnitude diagram for the Ultra Deep Field.  The error
bars are strictly due to number statistics and do not reflect completeness or systematic
errors.  For comparison the NHDF F814W and F160W (NHDF) data are overplotted.  The fall
off at the AB magnitudes greater than 27 is partially due to the conservative nature of the
source extraction used for the Treasury v2.0. catalog.}
\label{fig-mag}
\end{figure}

Although this is a description of the data and techniques utilized in 
constructing the NICMOS UDF Treasury Version 2.0 catalog, it has been 
traditional (\citet{wil96}, \citet{thm99}) to show a number magnitude 
diagram from the data. Figure~\ref{fig-mag} displays the number magnitude
diagram in AB magnitudes for the NICMOS F160W sources in the Ultra Deep Field.
The vertical error bars are calculated from number statistics only.  
Other errors such as incompleteness, source noise, and large scale structure 
are not included.  The fall off at AB magnitudes greater than 27 is 
certainly due to incompleteness, partially due to the conservative source 
extraction parameters used in the construction of the Treasury catalog.  
Note that the magnitudes are aperture magnitudes in the smallest, $0.6\arcsec$,
aperture.  For comparison the F814W data from the NHDF \citep{wil96} are
also plotted.  The agreement is good, with the F160W plot possibly having 
a slightly steeper slope, opposite to what was observed in the NHDF 
\citep{thm03}. 

\section{Acknowledgments}

We are extremely grateful to Massimo Stiavelli, Bahram Mobasher and Louis 
Bergeron for pointing out objects in our reduction that did not appear in 
their independent reduction of the same data and for providing one of us 
(RIT) with the images for detailed comparison. We are grateful to all of 
the personnel at STScI, and in particular Beth Perriello, who worked very 
hard to make the NICMOS UDF observations a success. This task was made even 
more difficult by the NCS safing that delayed the beginning of the
observations. This article is based on data from observations with the 
NASA/ESA Hubble Space Telescope, obtained at the Space Telescope Science 
Institute, which is operated by the Association of Universities for Research 
in Astronomy under NASA contract NAS 5-26555.  The individual researchers are 
funded in part by NASA Grant HST-GO-09803.01-A-G from the Space Telescope 
Science Institute. ST received support from the Danish Natural Research
Council.

\clearpage

\begin{deluxetable}{cccccccccccccccccccccccc}
\tabletypesize{\scriptsize}
\tablecaption{First Half Image Positions by Visit \label{tb-pos1}}
\tablewidth{0pt}
\tablehead{
\colhead{1} & \colhead{2} & \colhead{3} & \colhead{4} & \colhead{5} & \colhead{6} & \colhead{7} & \colhead{8} & \colhead{9} & \colhead{10} & \colhead{11} & \colhead{12} & \colhead{13} & \colhead{14} & \colhead{15} & \colhead{16} & \colhead{17} & \colhead{18} & \colhead{19} & \colhead{20} & \colhead{21} & \colhead{22} & \colhead{23} & \colhead{24}  
}
\startdata
B&C&T&B&C&T&B&C&T&B&C&T&B&C&T&B&C&T&B&C&T&B&C&T\\
\enddata

\end{deluxetable}

\begin{deluxetable}{cccccccccccccccccccccccc}
\tabletypesize{\scriptsize}
\tablecaption{Second Half Image Positions by Visit \label{tb-pos2}}
\tablewidth{0pt}
\tablehead{
\colhead{25} & \colhead{26} & \colhead{27} & \colhead{28} & \colhead{29} & \colhead{30} & \colhead{31} & \colhead{32} & \colhead{33} & \colhead{34} & \colhead{35} & \colhead{36} & \colhead{37} & \colhead{38} & \colhead{39} & \colhead{40} & \colhead{41} & \colhead{42} & \colhead{43} & \colhead{44} & \colhead{45} & \colhead{46} & \colhead{47} & \colhead{48}  
}
\startdata
R&C&L&R&R&L&R&C&L&R&C&L&R&C&L&R&L&C&L&R&R&L&R&R\\
\enddata

\end{deluxetable}

\begin{deluxetable}{ccccc}
\tabletypesize{\scriptsize}
\tablecaption{Data Quality Codes \label{tb-dqc}}
\tablewidth{0pt}
\tablehead{
\colhead{bit or number} & \colhead{Bad Pixel} & \colhead{Cosmic Ray} & \colhead{Non Linear} & \colhead{Saturated} 
}
\startdata
bit & 8 & 9 & 12 & 13 \\
number&256 & 512 & 4096 & 8192 \\
\enddata

\end{deluxetable}

\begin{deluxetable}{ccccccccccc}
\tabletypesize{\scriptsize}
\tablecaption{NICMOS Camera 3 distortion coefficients. \label{tb-dis}}
\tablewidth{0pt}
\tablehead{
\colhead{x or y} & \colhead{1} & \colhead{2} & \colhead{3} & \colhead{4} & \colhead{5} & \colhead{6} & \colhead{7} & \colhead{8} & \colhead{9} & \colhead{10} 
}
\startdata
x & 0. & 1.0014705 & 0. & 8.0317971E-6 & 1.3219373E-5 & 5.8285553E-6 & 0. & 0. & 0. & 0. \\
y & 0. & -8.9368516E-4 & 0.99853067 & -1.8073393E-5 & 0.59911861E-7 & -1.1582927E-5 & 0. & 0. & 0. & 0. \\
\enddata

\end{deluxetable}       

\clearpage

\begin{deluxetable}{cc}
\tabletypesize{\small}
\tablecaption{Images with special masks.  The numbers and letters are the unique sections of the file names. \label{tb-msk}}
\tablewidth{0pt}
\tablehead{
\colhead{F110W} & \colhead{F160W}
}
\startdata
01bl & 01ct\\
08me & 03jy\\
09tf & 03kj\\
13ap & 06et\\
18c3 & 07ka\\
20j6 & 13au\\
27es & 14d1\\
28im & 14dn\\
29jr & 16ve\\
30wj & 17b8\\
32xe & 20jb\\
33fh & 25c4\\
34m8 & 26cm\\
34mq & 27f0\\
35nt & 28jc\\
36ot & 28l1\\
38zd & 29l1\\
40qs & 31vp\\
44js & 32xl\\
45qg & 34my\\
\nodata & 35ny\\
\nodata & 36nx\\
\nodata & 36p0\\
\nodata & 38zk\\
\nodata & 39pe\\
\nodata & 40pw\\
\nodata & 42ht\\
\nodata & 43i7\\
\nodata & 44jz\\
\nodata & 45rl\\
\nodata & 48az\\
\enddata

\end{deluxetable}

\clearpage

\begin{deluxetable}{ccccccccc}
\tabletypesize{\small}
\tablecaption{The observed PSF parameters for the star in the NICMOS UDF images at (x,y)  (819.2,597.3), a calibration star P330-E drizzled in the same way as the UDF images, and a synthetic Tiny Tim image at the camera 3 focal position used in the UDF. \label{tab-psf}}
\tablewidth{0pt}
\tablehead{
\colhead{Parameter} & \colhead{F110W} & \colhead{F160W} & \colhead{F110W} & \colhead{F160W} & \colhead{F110W} & \colhead{F160W} & \colhead{F110W} & \colhead{F160W}\\
\colhead{} & \multicolumn{2}{c}{UDF Star} & \multicolumn{2}{c}{UDF Star\tablenotemark{a}} & \multicolumn{2}{c}{P330-E} & \multicolumn{2}{c}{Tiny Tim} 
}
\startdata
FWHM Major & $0.36\arcsec$ & $0.39\arcsec$ & $0.38\arcsec$ & $0.38\arcsec$ & $0.29\arcsec$ & $0.30\arcsec$ & $0.29\arcsec$ & $0.27\arcsec$\\
FWHM Minor & $0.32\arcsec$ & $0.36\arcsec$ & $0.35\arcsec$ & $0.35\arcsec$ & $0.26\arcsec$ & $0.26\arcsec$ & $0.25\arcsec$ & $0.24\arcsec$\\
\enddata
\tablenotetext{a}{These values were measured for the images kindly supplied by Stiavelli, Mobasher and Bergeron. See \S~\ref{ss-igi}}
\end{deluxetable}

\begin{deluxetable}{cccccc}
\tabletypesize{\scriptsize}
\tablecaption{Source extraction parameters use in the NICMOS Treasury version 2.0 catalog.  Obvious parameters such as DETECT\_TYPE = CCD have not been entered.  The full configuration files are part of the version 2.0 Treasury archive in MAST. rmsfilter.fit refers to the rms image for a given filter such as rmsF160W.fit. \label{tb-separ}}
\tablewidth{0pt}
\tablehead{
\colhead{Parameter} & \colhead{Value} & \colhead{Parameter} & \colhead{Value} & \colhead{Parameter} &\colhead{Value}
}
\startdata
THRESH\_TYPE & RELATIVE & DETECT\_MINAREA & 7 & DETECT\_THRESH & 1.4\\
ANALYSIS\_THRESH & 1.4 & FILTER & Y & FILTER\_NAME & detec.conv\\
DEBLEND\_NTHRESH & 32 & DEBLEND\_MINCONT & 0.001 & CLEAN & Y\\
CLEAN\_PARAM & 5 & MASK\_TYPE & CORRECT & WEIGHT\_IMAGE & rms\_combfix.fit,rmsfilter.fit\\
WEIGHT\_TYPE & RMS\_WEIGHT & PHOT\_APERTURES & 6,11,17 & PHOT\_AUTOPARAMS & 2.5,3.5\\
\enddata

\end{deluxetable}

\clearpage

\begin{deluxetable}{cccccccccccc}
\tabletypesize{\scriptsize}
\rotate
\tablecaption{A mini catalog of source parameters. See \S~\ref{ss-mc} for a desription of the parameters. \label{tab-cat}}
\tablewidth{0pt}
\tablehead{
\colhead{ID} & \colhead{ACS} & \colhead{SEG} & \colhead{X} & \colhead{Y} & \colhead{RA}
& \colhead{DEC} & \colhead{PA} & \colhead{Elip} & \colhead{Rh} & \colhead{FWHM} 
& \colhead{stel}
}
\startdata
   1& 4394& 605&2877.54&1797.00& 53.13059235&-27.79026222&  0.0000&  0.8470&  2.9450&  4.3800&0.840\\
   2& 5109& 631&2789.20&1766.72& 53.13309097&-27.79101944&-23.9000&  0.1540&  2.0870&  7.1800&0.890\\
   3& 5010& 636&2764.20&1762.93& 53.13379669&-27.79111481& 78.0000&  0.5340&  4.0370&  8.7800&0.910\\
   4& 4638& 702&2750.66&1709.68& 53.13417816&-27.79244423&-45.3000&  0.4250&  2.8050&  5.6200&0.700\\
   5& 3460& 926&2697.05&1500.98& 53.13569260&-27.79766273& 45.3000&  0.4080&  3.3700&  4.1400&0.730\\
   6& 5115& 627&2685.79&1769.54& 53.13601303&-27.79094887& 27.4000&  0.3890&  2.9230&  7.2000&0.920\\
   7& 3871& 842&2673.44&1586.07& 53.13636017&-27.79553413& 68.7000&  0.4570&  3.9490&  9.2400&0.520\\
   8& 4350& 774&2671.81&1643.54& 53.13640594&-27.79409790&-21.0000&  0.0550&  2.8130&  5.1000&0.980\\
   9& 3938& 867&2666.78&1572.05& 53.13655090&-27.79588509& 17.7000&  0.0380&  2.5200&  4.1100&0.940\\
  10& 4370& 792&2661.00&1637.50& 53.13671112&-27.79424858& 89.8000&  0.1730&  2.7700&  4.6100&0.430\\
\enddata
\end{deluxetable}

\clearpage

\clearpage

\clearpage

\clearpage

\end{document}